\documentclass[a4paper,fleqn]{cas-sc}

\usepackage[numbers]{natbib}
\usepackage{ulem}

\usepackage{lineno}

\def\tsc#1{\csdef{#1}{\textsc{\lowercase{#1}}\xspace}}
\tsc{WGM}
\tsc{QE}
\tsc{EP}
\tsc{PMS}
\tsc{BEC}
\tsc{DE}

\begin{document}
	\let\WriteBookmarks\relax
	\def\floatpagepagefraction{1}
	\def\textpagefraction{.001}
	\shorttitle{Uniform microcapsule production by emulsification }
	\shortauthors{Maleki et~al.}
	
	\title [mode = title]{Membrane emulsification for the production of suspensions of uniform microcapsules with tunable mechanical properties}

	\author[1]{Mehdi Maleki}
	\author[1]{Cl\'ement de Loubens}[orcid=0000-0002-4988-9168]
	\cormark[1]
	\author[1]{Kaili Xie}
	\author[1]{Emeline Talansier}
	\author[1]{Hugues Bodiguel}
	\author[1]{Marc Leonetti}
	
	\address[1]{Univ. Grenoble Alpes, CNRS, Grenoble INP, LRP, 38000 Grenoble, France}

	\cortext[cor1]{Corresponding author:  clement.de-loubens@univ-grenoble-alpes.fr}
	
	\begin{abstract}
		A way forward for high throughput fabrication of microcapsules with uniform size and mechanical properties was reported irrespective of the kinetic process of shell assembly. Microcapsules were produced using lab-scale emulsification equipment with a micro-engineered membrane in the size range 10-100 $\mu$m. The shell of the microcapsules was assembled  at the water-oil interface by complexation of polyelectrolytes or cross-linking of proteins providing two different kinetic processes. Elasticity of microcapsules was characterized with an automated extensional flow chamber. Process parameters were optimized to obtain suspensions with size variations of 15\%. Some strategies were developed to obtain uniform elastic properties according to the kinetics of shell assembly. If kinetics is limited by diffusion, membrane emulsification and shell assembly have to be split into two steps. If kinetics is limited by the quantity of reactants encapsulated in the droplet, variations of elastic properties result only from size variations.
	\end{abstract}

	\begin{highlights}
		\item Fabrication of microcapsules by membrane emulsification
		\item Automatic measurement of the elasticity of microcapsules
		\item High throughput production of uniform microcapsules of size 10-100 $\mu$m.
		\item Different strategies to control shell properties and uniformity
	\end{highlights}
	

	\maketitle

\section{Introduction}

Tuning the physical properties of microcapsules is of prime importance to control their stability and the delivery of encapsulated compounds in biological or industrial  processes \cite{Anandhakumar2010, McClements2015, trojer2015use, Neubauer:2014aa}. Microcapsules' shells are ultra-thin elastic films characterized by a surface shear elastic modulus $G_s \sim Gh$, where $h$ is the shell thickness and $G$ the bulk shear elastic modulus. Deformation of a microcapsule of radius $R$ in a shear flow of hydrodynamic stress $\sigma$ is controlled by a capillary number $Ca = \sigma R/G_s$ that represents the ratio of viscous stress over the elastic shell response \cite{ChangKS1993}. Consequently both size and shell rheology  control the fate of microcapsules in processes such as break-up. Release of encapsulated compounds by osmotic pressure differences is also controlled by both shell elasticity and  capsule size \cite{gao2001elasticity}.

The simplest process for the fabrication of microcapsules is based on the emulsification with a rotor of two immiscible fluids with chemicals in each phase and their subsequent reaction at the interface leading to the formation of the shell \cite{Clement2014, xie2017interfacial}. The major drawback of this process is the large  variations of size and elasticity of produced microcapsules \cite{Clement2014, gubspun2016characterization}. Up to now, the gold standard method to control the properties of microcapsules  is the layer-by-layer assembly of polyelectrolytes on dissolvable particles \cite{richardson2016innovation}. However this method is long and complex if several layers have to be deposited in order to stiffen the shell. Recently, several research groups have developed microfluidics techniques to generate microcapsules with highly uniform size and elasticity based on flow-focusing systems \cite{xie2017interfacial, sivakumar2008monodisperse,hennequin2009synthesizing, kaufman2015soft, de2017one}. These techniques have clearly demonstrated their usefulness in understanding the relationships between the assembly of microcapsules and their interfacial rheological properties \cite{xie2017interfacial, tregouet2018microfluidic}, but were limited to  research tools as their throughput was very low (a few $\mu$L/min) and were not easily scalable  for industrial applications.  Designing high throughput processes for the fabrication of microcapsule suspensions with uniform physical properties should open new perspectives for the optimization of microencapsulation technologies, e.g. drug delivery systems. 

On the other hand, producing high-quality emulsions at pilot and industrial scales was made possible by the development of membrane emulsification systems with micro-engineered membranes \cite{joscelyne2000membrane, vladisavljevic2012production, doi:10.1021/ie0504699}. Briefly, membrane emulsification consists in passing the dispersed phase through a membrane made of regular array of pores to form droplets that are detached by hydrodynamic stress. The pores size and surface properties have to be chosen carefully in order to produce monodisperse drops \cite{CHRISTOV200283}. To some extent, membrane emulsification is a scaled-up version of  microfluidic T-junction chips. Using this technique, suspensions of particles \cite{IMBROGNO2015116}, multiple emulsions \cite{VLADISAVLJEVIC201478}, soft beads \cite{HANGA20141664} and liposomes \cite{VLADISAVLJEVIC2014168} were produced at lab and pilot scales with an excellent control of the size uniformity.

However, translating membrane emulsification technologies for the high throughput production of microcapsules is not straightforward to obtain particles with uniform size and elastic properties. In fact, shell elasticity depends closely on the physico-chemical conditions of shell formation \cite{fery2007mechanical, dubreuil2003elastic, A.Walter2000}, but also on the control parameters of the fabrication process \cite{Clement2014, xie2017interfacial}. Two  well-studied examples can illustrate the differences in the kinetics of shell formation. The first one is based on interfacial complexation of polyelectrolytes. Two oppositely charged polyelectrolytes are dissolved in two immiscible phases and form a solid  at the interface of the droplets  after the emulsification step.  It has been established that the kinetics of shell formation is limited by the diffusion of one of the polyelectrolytes \cite{grigoriev2008new, Deniz2011,sivakumar2008monodisperse, rinaudo2008surfactant, xie2017interfacial}. Consequently, surface elasticity increases with the  shell formation time (from a few seconds to a few hours) \cite{xie2017interfacial}. The second example relates to microcapsules based on interfacial cross-linking of proteins  \cite{Levy1996, lu2004microcapsule}. In this case, proteins are dissolved in an aqueous phase and dispersed in an organic phase that contains the cross-linker. The reaction is limited to a few seconds by the quantity of proteins encapsulated in the droplet and surface elasticity greatly increases with the capsule size, whereas other parameters are kept constant (i.e. pH, concentration of proteins) \cite{Clement2014,gubspun2016characterization,xie-thesis}. Conversely, the surface elasticity of polyelectrolytes microcapsules does not depend on their size \cite{xie2017interfacial}.   Both examples show that time and size have different impacts on the mechanical properties of the shell depending on its kinetics of formation. We can also anticipate that the membrane emulsification process should be controlled in different ways  based on the kinetics of shell formation if elastic properties have to be controlled. The surface elasticity of the shell can be characterized by several methods: osmotic tests \cite{gao2001elasticity}, AFM \cite{fery2004mechanics} or hydrodynamic methods \cite{Clement2014}. To date, these techniques are manual and time consuming. Consequently, a small number of capsules can be analyzed. In order to quantify the shell elasticty and its variation in a sample and to conclude about the performance of membrane emulsification for capsule production, it is so required to develop also automated systems of capsule characterization.

Our objective was to show a route for high throughput fabrication of microcapsules with uniform size and elasticity. Suspensions of microcapsules were produced with a scalable membrane emulsification system   \cite{doi:10.1021/ie0504699} using a micro-engineered membrane in the size range of 10-100 $\mu$m. The effect of  process parameters on the size distribution was studied.  In order to optimize the process according to the kinetics of shell formation, two well-characterized microencapsulation systems  were studied based on interfacial complexation of polyelectrolytes \cite{Deniz2011, xie2017interfacial} and cross-linking of a protein \cite{Clement2014,Clement2015, gubspun2016characterization}.  Based on a well-established hydrodynamic method \cite{Clement2014}, we developed a new millifluidic chamber to measure at high-throughput the distribution of the surface elasticity of a suspension. In this way, we designed different strategies to optimize the uniformity of the elastic properties according to the kinetics of shell formation.
	
\begin{figure}[t]
	\centering
	\includegraphics[width=0.7\textwidth]{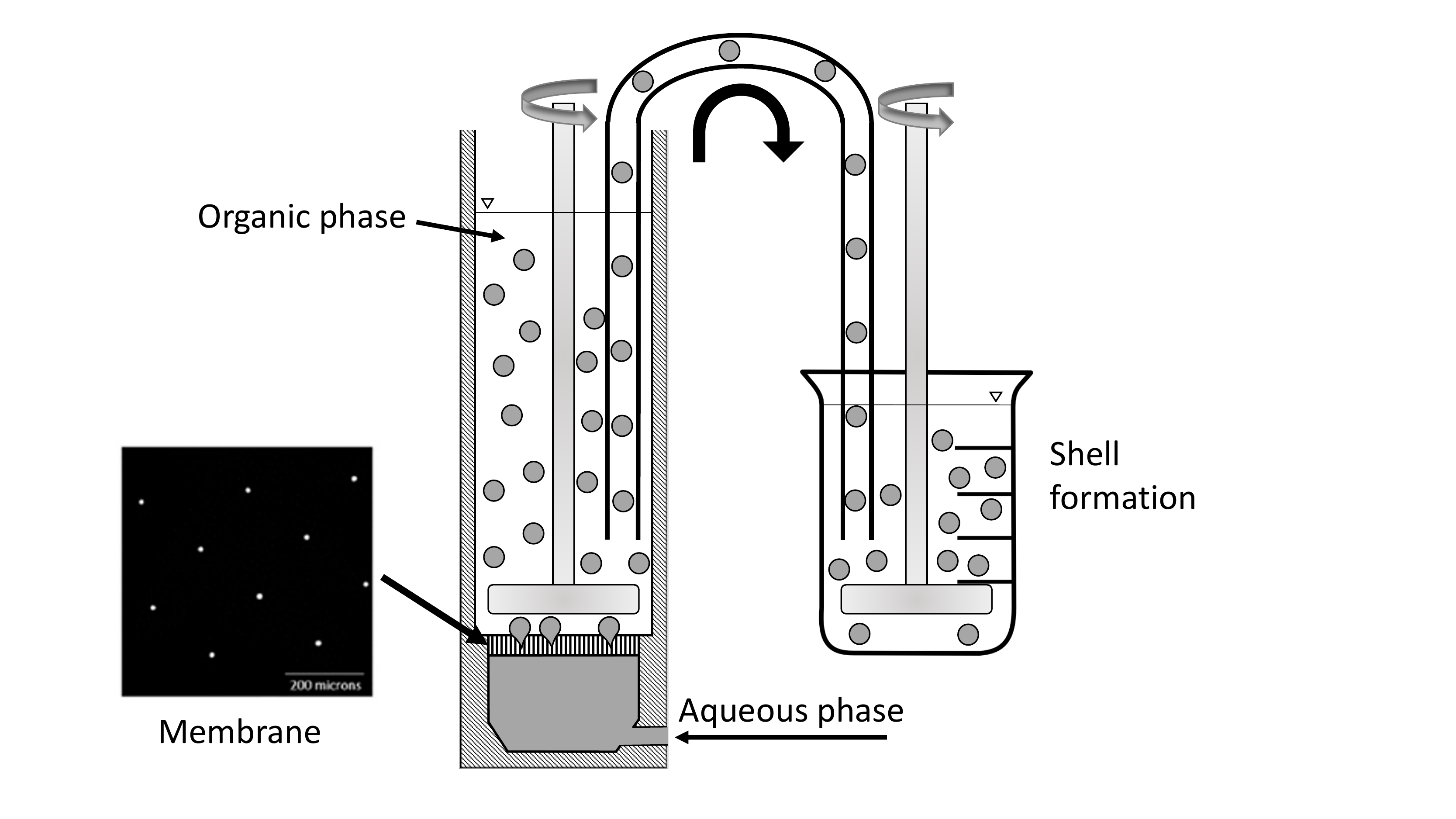}
	\caption{ Sketch of the membrane emulsification system  and microscopic image of the micro-engineered membrane. Aqueous droplets of BSA or CH were dispersed  in the oil phase through the micro-engineered membrane. The droplets were sucked-up in a  beaker that contained TC or PFacid and stirred for several minutes in order to build the BSA/TC or CH/PFacid shell. 
	}
	\label{methods}
	
\end{figure}

\section{Materials and Methods}
\subsection{Solutions for microcapsule preparation}

Microcapsules were made by interfacial complexation of chitosan (CH) with phosphatidic fatty acid (PFacid) \cite{Deniz2011, xie2017interfacial} or by interfacial cross-linking of bovine serum albumin (BSA) with terephthaloyl chloride (TC)  \cite{Edwards_L_vy_1993}. 

Chitosan is a polysaccharide carrying positively charged groups and is soluble in water at pH 3.0, whereas PFacid is a surfactant that is soluble in vegetable oil and negatively charged. Chitosan (CAS number 9012-76-4, Sigma-Aldrich) of medium molecular weight and 75-85\% deacetylation was dissolved in Millipore water  (Resistivity > 18.2 $m\Omega .cm$) at a concentration of 0.25 \% w/w by adjusting the pH with hydrochloric acid (1 mol/L) at 3.0. The chitosan solution was then filtered to remove undissolved particles through Minisart syringe-filters (pore size 5.0 $\mu m$). PFacid was  comprised of a commercial lecithin known as lecithin YN (Palsgaard 4455, food-grade, E442) which was kindly provided   by Palsgaard. PFacid was dissolved in food-grade colza oil (G\'eant Casino, french supermarket) with concentration ranging from 0.1 to 10 \% w/w.

To synthesize BSA / TC  microcapsules, BSA (CAS number  9048-46-8, Sigma-Aldrich) was  dissolved in phosphate buffered saline solution (Fisher Scientific) and TC (CAS number 100-20-9, Sigma-Aldrich) dissolved in a mixture of chloroform (CAS number 67-66-3, Sigma-Aldrich) : cyclohexane (CAS number110-82-7, anhydrous, 99.5\%, Sigma-Aldrich) at a ratio of 4:1 by volume. Sorbitane trioleate 85 (SPAN 85, CAS Number: 26266-58-0, Sigma-Aldrich) was used to stabilize albumin drops during emulsification.

\subsection{Fabrication of  suspensions of microcapsules}
Suspensions of microcapsules were prepared with a lab-scale membrane emulsification system marketed by Micropore Technologies Ltd (UK) under the commercial name Micropore LDC-1. As shown in Figure \ref{methods}, it included a micro-engineered emulsification membrane under a  paddle-blade stirrer. The stirrer was driven by a DC motor, which controlled the rotational velocity $\omega$. The  emulsification membrane was a thin flat nickel membrane which was chemically treated to have an hydrophobic surface. It was provided by Micropore Technologies Ltd (UK). The cylindrical pores were made by laser beam with a diameter of 10  $\mu$m. The pore spacing and porosity of the membrane were 200 $\mu$m and 0.23 \%. The array of the pores was located in an annular narrow ring shape region on the membrane to limit variation of shear rate and minimize the polydispersity \cite{doi:10.1021/ie0504699}. 

For CH / PFacid microcapsules, PFacid solution was agitated by the paddle with a rotational speed of 400 to 1200 rpm. The chitosan solution was injected for 45 s by a syringe pump (Nemesys) through the membrane. The droplets were sucked-up from the dispersion cell with a tube and collected in a beaker to improve the size uniformity of the suspension (see Results and Discussion). The capsules were kept under gentle stirring in the PFacid solution for 2 to 30 min to stiffen the shell.  
For BSA / TC microcapsules, solution of vegetable oil with span 85 was agitated by the paddle with a rotational speed of 400 to 1200 rpm. The BSA solution was injected by a syringe pump  through the membrane during 45 s.  The droplets were sucked-up from the dispersion cell and collected in a beaker containing the solution of TC. The capsules were kept under gentle stirring in the beaker for 2 min to stiffen the shell.  The mean size of the capsules ranged from 10 to 100 $\mu$m, according to the process parameters and viscosities of the dispersed and continuous phases.

For both encapsulation systems, the shell growth was stopped by diluting the oil phase with a large quantity of cyclohexane. The microcapsules were then suspended in silicon oil AP1000 (Wacker) for subsequent analysis.

\begin{figure}[t]
	\centering
	\includegraphics[width=0.75\textwidth]{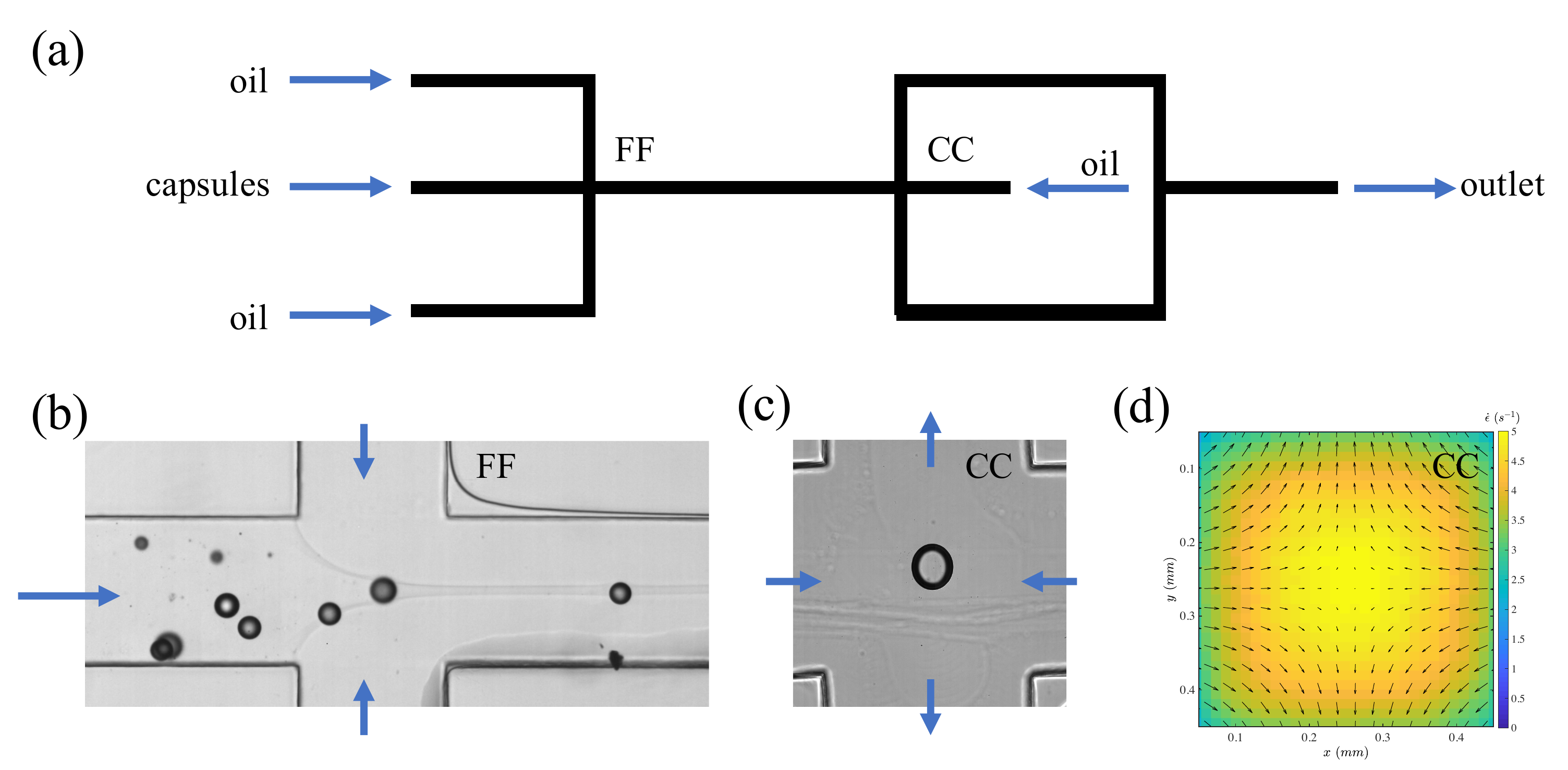}
	\caption{Automated extensional flow chamber for the characterization of microcapsules.  \textbf{a:} Design of the millifluidic device. Capsules and oil are injected in a flow focusing geometry (FF) to center the capsules in the channel. Capsules are then stretched in the center of the cross-slot channel (CC) .\textbf{b:}  Microcapsules which have been produced by membrane emulsification being injected in the FF geometry. \textbf{c:}  Deformation of a capsule in the center of the cross by extensional flow, see \ref{elongation} for details. \textbf{d:} Colormap of the rate of elongation $\dot\epsilon$ in the center of the cross junction. 
	}
	\label{elongationSetUp}
	
\end{figure}

\subsection{Characterization of  microcapsules}\label{elongation}
Size distribution of microcapsules was measured by bright field microscopy. A diluted suspensions of microcapsules was let to sediment in a petri dish and scanned by an inverted microscope Olympus IX-72 equipped with a 4, 10 or 20 times lens, a Marzhauser motorized stage and a camera Hamamatsu ORCA-Flash 4.0. Consequently, microcapsules were all on the same focal plane without overlap between them. The size of the microcapsules was measured with a home-made algorithm based on the image processing toolbox of Matlab (Mathworks). Due to the difference of optical indexes between oil and water, the contrast was excellent which make the detection not sensitive to the threshold or the lighting conditions. The measurements were carried out on several thousand capsules for each batch. The capsules were deposited on a glass slice (in a very diluted regime), so they were all on the same plane without overlap between them. Typically, the mean radius ranged from 10 to 100 µm. The standard deviation normalized  by the mean radius ranged from  15 to 50\%, see Results for details.

The surface shear elasticity $G_s$ of the microcapsules was measured in an extensional flow chamber that has been automatized to study a large number of microcapsules. In our previous work, the determination of $G_s$ took 5 min for a single capsule \cite{Clement2014,xie2017interfacial,Clement2015} . Here, we analyzed about 1,000 microcapsules in 90 min.  Briefly, the device was comprised of a 2D flow-focusing geometry in serial with a cross-slot chamber, Figure \ref{elongationSetUp}-a. The flow-focusing geometry was used to center the capsules in the mid-plan of the channel. The cross-slot with two opposite inlets and two opposite outlets was used to generate an extensional flow in which microcapsules were deformed and analyzed. The channels of square section (1 $\times$ 1  mm$^2$) were engraved in PMMA and connected to four syringe pumps (Nemesys).  Microcapsules were injected into the central channel of the flow focusing and were centered by the two perpendicular jets (Figure \ref{elongationSetUp}-b). They then reached the center of the cross-slot chamber to be deformed in the region where the hydrodynamic stress was maximal (Figure \ref{elongationSetUp}-c, d). The velocity $\Vec{u}$ and the elongation rate $\dot{\epsilon}$ were calibrated by particle tracking velocimetry. As expected, we found that the flow was hyperbolic: $u_x=\dot{\epsilon}x$, $u_y=\dot{\epsilon}y$ with $\dot{\epsilon}$  being the rate of elongation (Figure \ref{methods}-c). $\dot{\epsilon}$ was almost constant in a zone of 200 $\mu$m radius around the stagnation point and increased linearly with the flow rate, see \cite{Clement2014} for details. The radius of this zone has to be at least 1.5 times larger than the radius of the microcapsules to obtain an accurate result \cite{Clement2014}. As the flow had a parabolic profile in the $z$ direction, we discarded from the analysis the microcapsules that were not localized in a region near the mid-plane of the channel using a homemade algorithm. Microcapsules deformed as an ellipsoid with a steady-state shape. The deformation is characterized by the Taylor parameter $D=(L-S)/(L+S)$ with $L$ and $S$ being the the major and minor axis of the ellipsoid respectively. In the regime of small deformations ($D_\infty$<0.1), the surface elasticity $G_s$ and the steady-state Taylor parameter $D_\infty$ are related by \cite{DBB1985}

\begin{equation}
	G_s = \frac{25}{6}\frac{\eta \dot{\epsilon} R}{D_\infty}
\end{equation}

with $\eta$ being the viscosity of the continuous phase, i.e. the silicone oil, 1.18 Pa.s at 22$^{\circ}$C.  

Results on the distribution of the size and the elasticity were presented in terms of probability distribution function. If $X$ is a variable of interest ($R$ or $G_s$), its probability distribution function  $P_X(X)$ is defined by

\begin{equation}
	P_X(X) = \frac{1}{\Delta X} \frac{N(X- \frac{\Delta X}{2},X+ \frac{\Delta X}{2} )}{N_T}
\end{equation}

where N is the number of microcapsules whose variable $X_i$ ranged between $X- \frac{\Delta X}{2}$ and $X+ \frac{\Delta X}{2}$ and $N_T$ the total number of microcapsules, so that $\int P_X dX = 1$. $P_X$ was then smoothed with an averaged moving filter. The uniformity was quantified  by the coefficient of variation $CV$

\begin{equation}
	CV = \frac{1}{\bar{X}}\sqrt{\frac{\sum_{i} (X_i-\bar{X})^2}{N_T-1}}\times 100
\end{equation}

with $\bar{X}$ being the mean value of X.

\begin{figure}[t]
	\centering
	\includegraphics[width=0.85\textwidth]{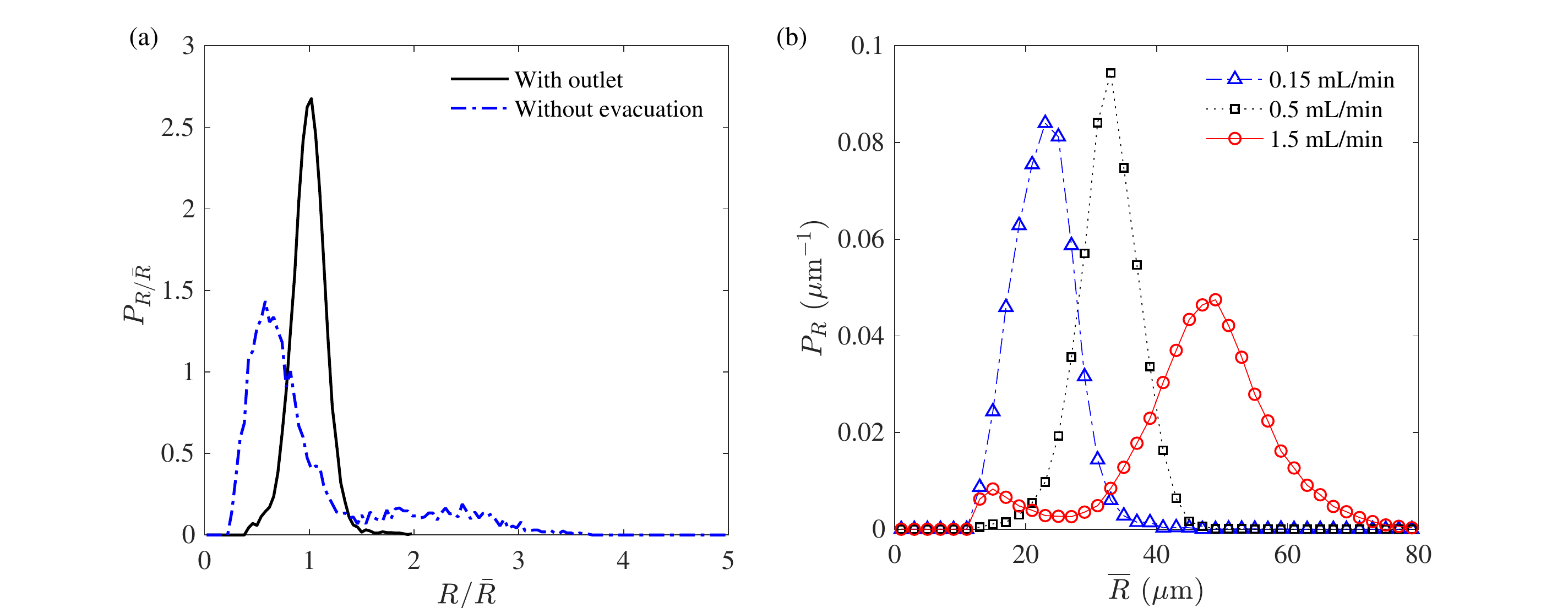}
	\caption{\textbf{a:} Probability distribution function of the size $P_R$ for suspensions of CH / PFacid microcapsules made with (plain line) and without (dashed line) outlet. \textbf{b:}  Comparison of $P_R$ for different flow rates $Q$ of CH solution injection. }
	\label{pdfR}
	
	\hspace{1cm}
	
	\centering
	\includegraphics[scale=0.35]{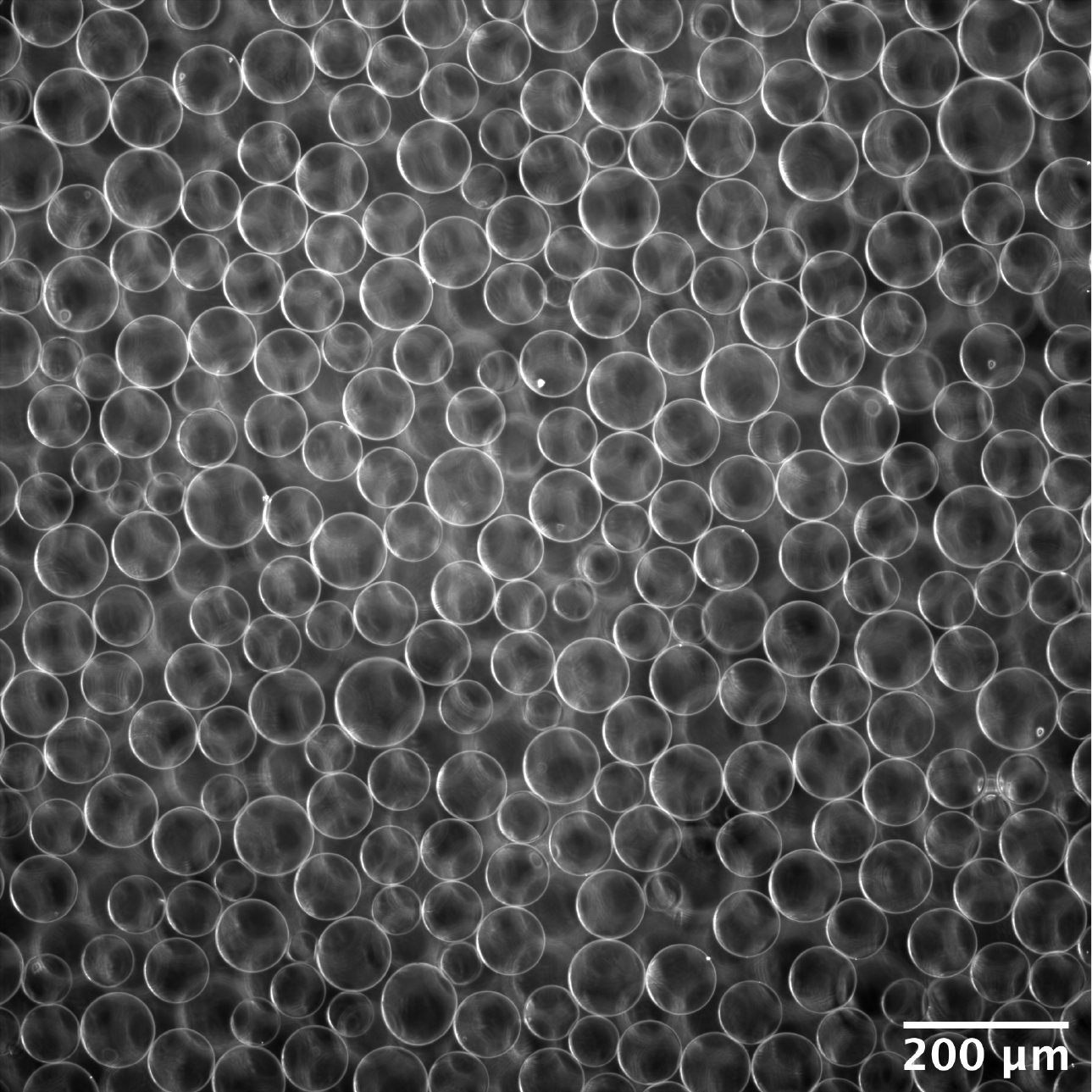}
	\caption{Fluorescence imaging of a suspension of CH / PFacid microcapsules produced by membrane emulsification.}
	\label{photo}
\end{figure}

\begin{figure}[t]
	\centering
	\includegraphics[width=0.7\textwidth]{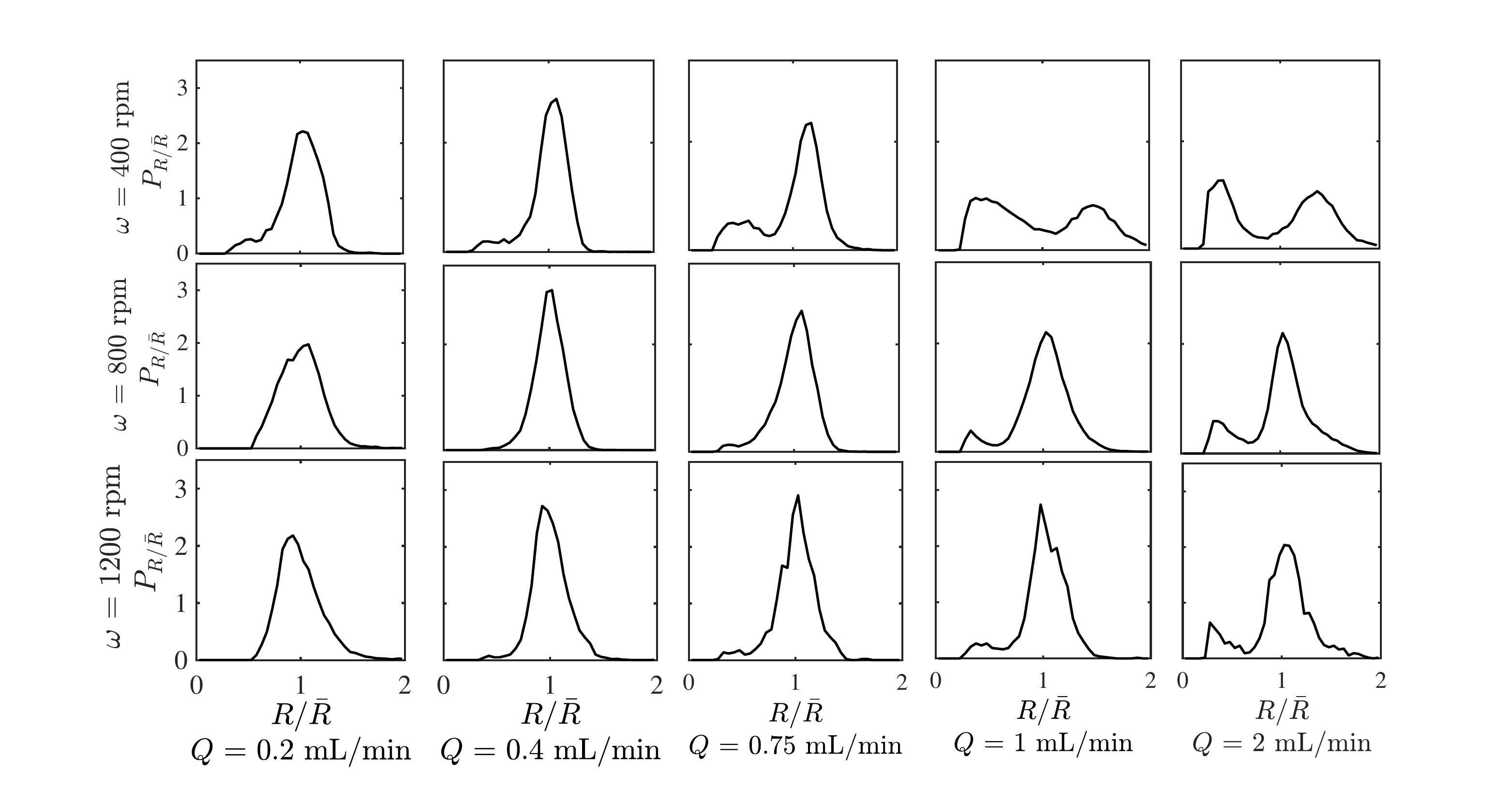}
	\caption{Size probability distribution $P_{R/\bar{R}}$ as a function of the normalized radius $R/\bar{R}$ for different paddle rotational speeds $\omega$ and injection flow rate $Q$. CH / PFacid microcapsules.}
	\label{distribution}
	
	\hspace{1cm}
	
	\centering
	\includegraphics[width=1\textwidth]{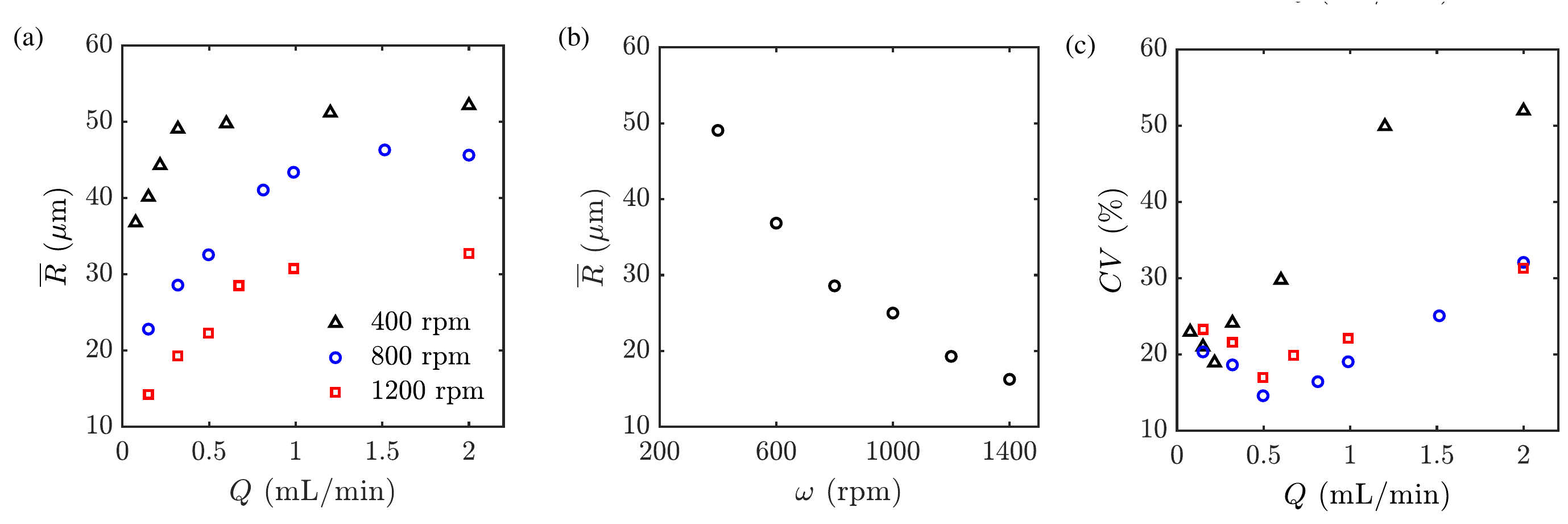}
	\caption{\textbf{a:} Mean radius $\bar{R}$ as a function of the injection flow rate $Q$ for different paddle rotational speed $\omega$. \textbf{b:} $\bar{R}$ as a function of $\omega$, $Q=$0.32 mL/min. \textbf{c:} Coefficient variation  of size $CV$ as a function of $Q$.  CH / PFacid microcapsules.}
	\label{optimum}
\end{figure}

\section{Results and Discussion}

\subsection{Size distribution of microcapsules}
We investigated the effects of process parameters (paddle rotational speed $\omega$ and flow rate $Q$) on the mean radius  $\bar{R}$, the probability distribution $P_R$ and coefficient of variation $CV$ for CH / PFacid microcapsules. First of all, the stirred emulsification cell was used as a batch process  to produce monosized particles \cite{doi:10.1021/ie0504699, egidi2008membrane}. For microcapsules, we observed that this batch version of the process was not suitable: $CV$ was larger than 50 \% (Figure \ref{pdfR}-a, dashed line). As the dispersed phase had a larger density than the vegetable oil, microcapsules were not efficiently suspended in the cell by the paddle and tended to settle near the membrane where they could coalesce and/or be broken by the mechanical agitation. To overcome this limitation, microcapsules were pumped continuously at constant rate from their site of formation to outside through a tube. As shown on Figure \ref{methods}, this strategy, similar to a continuous process, allowed us to minimize $CV$ up to 15\% in optimal conditions (Figure \ref{pdfR}-a; plain line).  Even with evacuation,  bi-modal  size distribution was observed  for large injection flow rates ($Q=$1.5 mL/min), whereas for lower $Q$ the distribution was Gaussian with a $CV$ that was minimized up to 15\% (Figure \ref{pdfR}-b and \ref{photo}).

Figure \ref{distribution} shows the shape of $P_R$ for systematic variations of $Q$ and $\omega$ . We observed that, by increasing $Q$, the shape of $P_R$ shifted from unimodal to bimodal form. This transition corresponded to the transition of drop generation from dripping (drop by drop) to jetting (continuous jet) regime \cite{dripping-bertrandias,dripping-meyer}. In the dripping regime (small $Q$), droplet detachment depends on the balance between interfacial force and hydrodynamic stress generated by the stirrer $\omega$.  It has been shown that droplets mature at the pore in a reproducible way until they are detached by hydrodynamic stress. This mechanism leads to a unimodal shape of $P_R$ \cite{dripping-bertrandias,dripping-meyer}. In the jetting regime (large $Q$), the inertial force of injection exceeds the interfacial force and droplets are created by a jet. The resulting jet breaks up by the Plateau-Rayleigh instability and alternates production of small and large droplets which leads to a bimodal shape of $P_R$  \cite{dripping-bertrandias,dripping-meyer}. Size distributions of Figure \ref{distribution} are summed up in Figure \ref{optimum} by plotting the variations of $\bar{R}$ and $CV$ with $Q$ and  $\omega$. Figures \ref{optimum}-a \& b show that $\bar{R}$ increased when $Q$ was increased and $\omega$ was decreased. At high $Q$ (>1 mL/min), $\bar{R}$ saturated due to the emergence of a bimodal size distribution (Figure \ref{distribution}).  Figures \ref{optimum}-c  show that $CV$ had an optimum value of 15 \% for 800 rpm and 0.5 mL/min. For other values of $Q$ and $\omega$, $CV$ was reasonably maintained at 20-25 \%, except for bimodal distribution, for which $CV$ exceeded 50 \%. This was coherent with previous results on droplets produced with a similar device \cite{doi:10.1021/ie0504699, egidi2008membrane}. Note that $CV$ was minimized up to few percents by using rotating membrane
	emulsification systems \cite{VLADISAVLJEVIC2014168}.

\begin{figure}[pht!]

		\centering
	\includegraphics[width=0.8\textwidth]{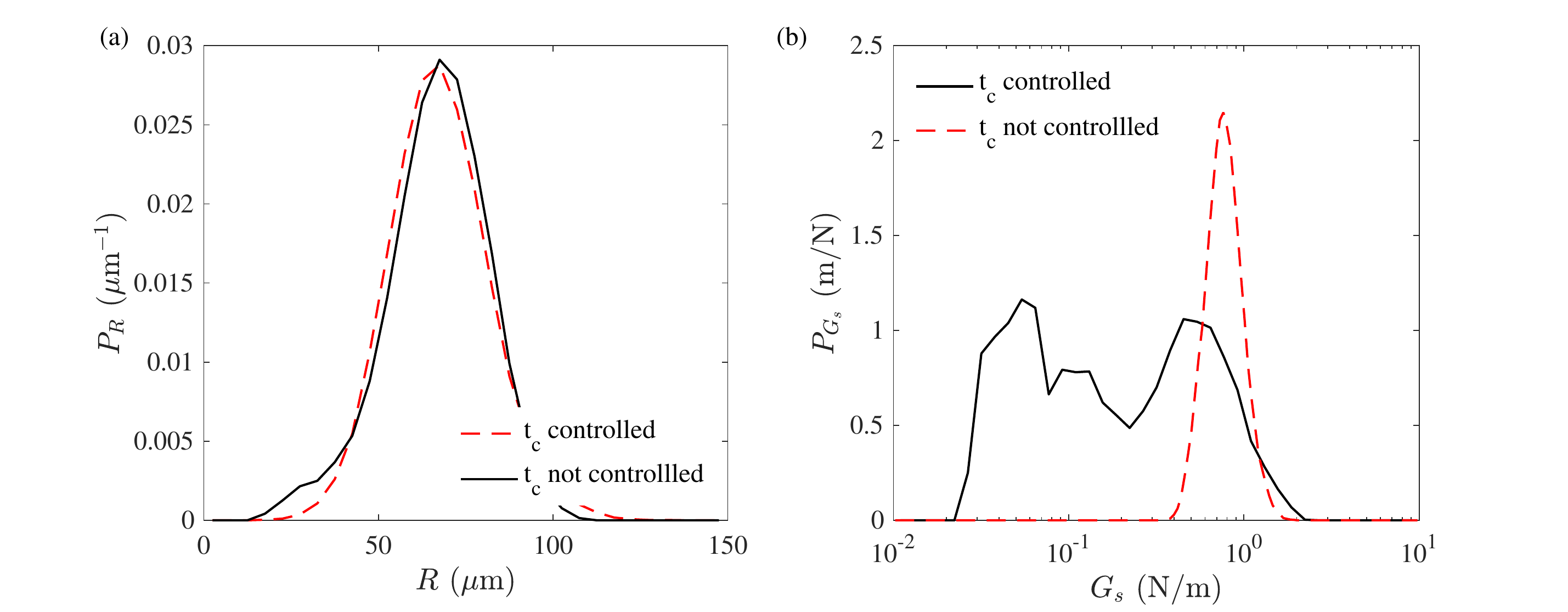}
	\caption{Comparison of uniformity of CH/PFacid microcapsules with (red dashed  line) and without (black plain line) control of complexation time $t_c$. \textbf{a:}  Probability distribution function of size $P_{R}$. $CV$ is of 18 -19\% for both conditions.   \textbf{b:} Probability distribution function of surface elasticity $P_{G_s}$. $CV$ is of 56\% when the $t_c$ is not controlled and reduced up to 24\% when $t_c$ is controlled.}
	\label{pdfGs_CH2}
	
	\centering
	\includegraphics[width=0.75\textwidth]{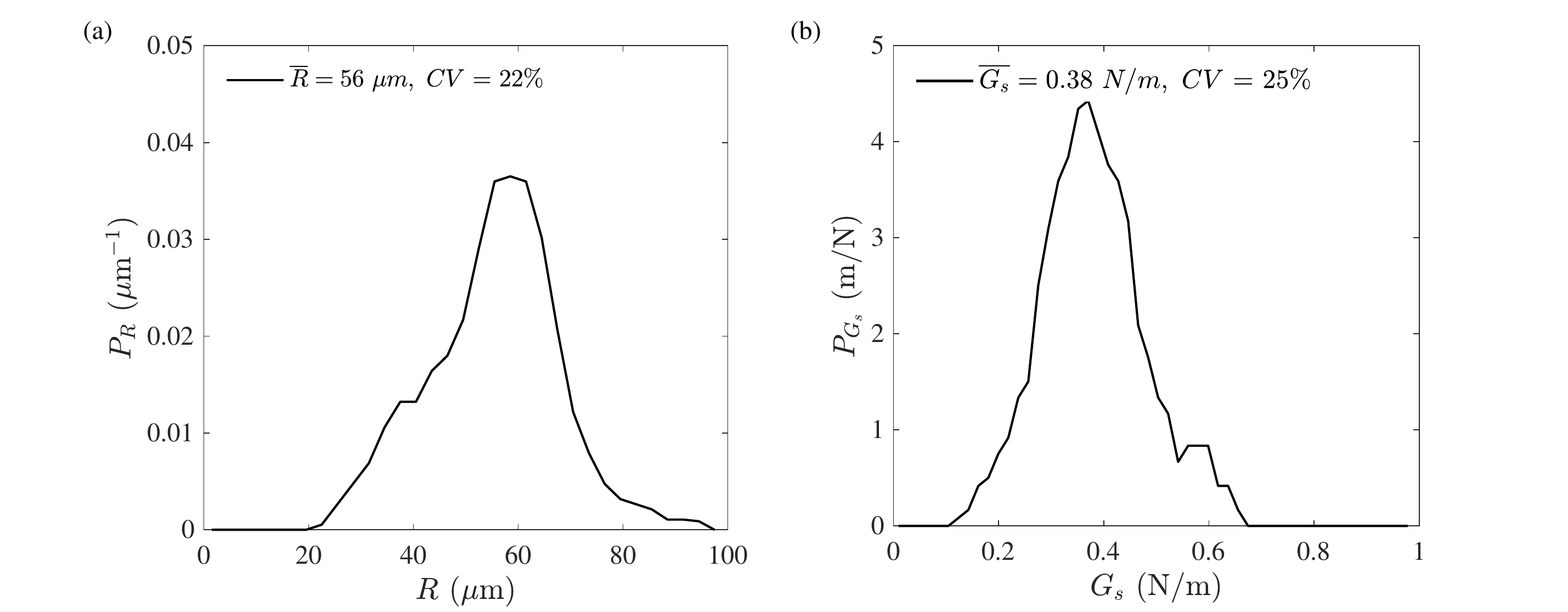}
	\caption{Fabrication and characterization of BSA/TC microcapsules. \textbf{a:} Size probability distribution $P_R$. \textbf{b:}  Surface elasticity distribution $P_{G_s}$.}
	\label{BSA}
	
		\hspace{1cm}
		
		\centering
	\includegraphics[width=0.5\textwidth]{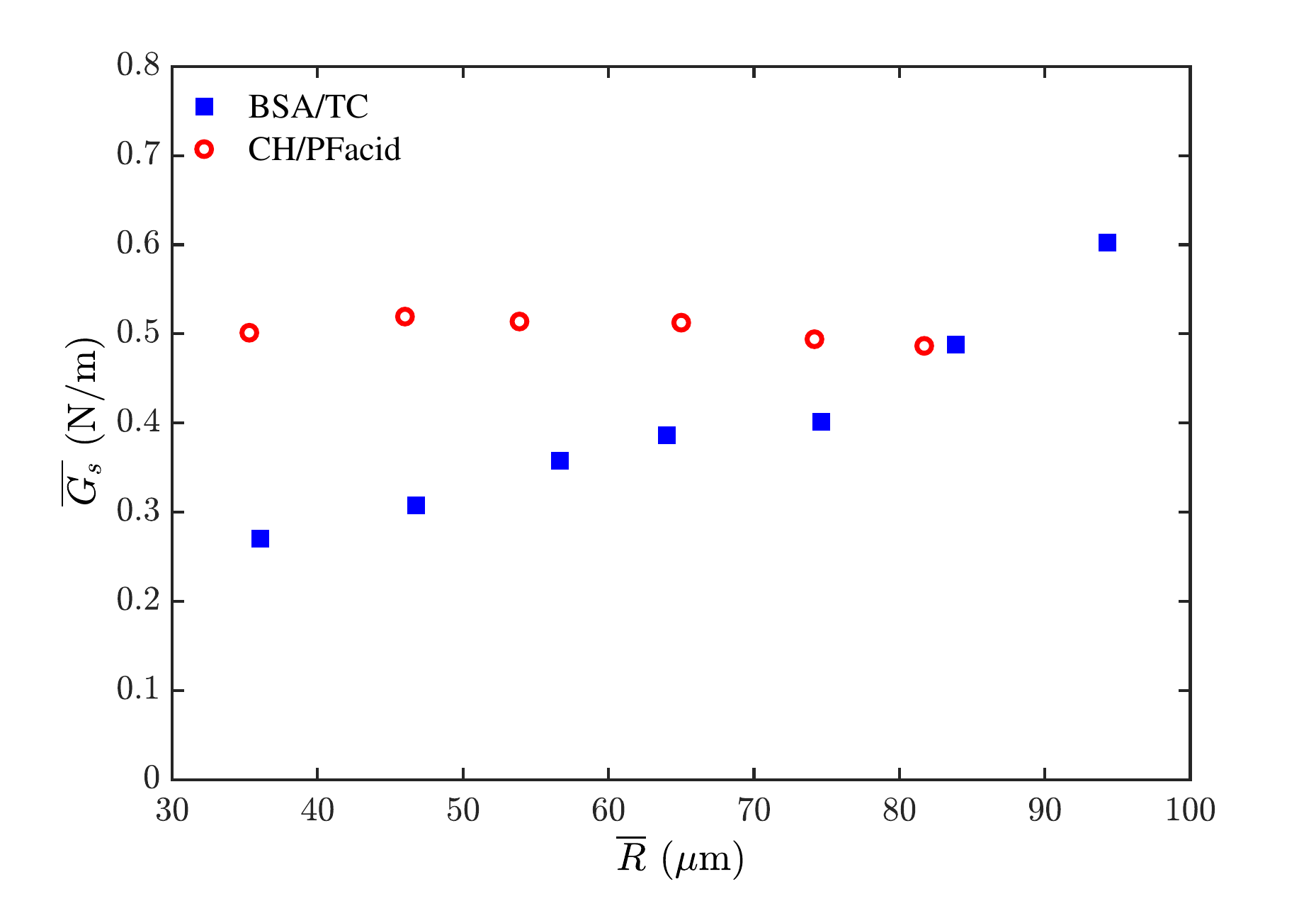}
	\caption{Mean surface elasticity $\bar{G_s}$ as a function of the size range ($\pm$ 10 $\mu$m).}
	\label{BSACHsize}
\end{figure}

\subsection{Microcapsules with uniform elasticity}

During production of uniform droplets by membrane emulsification, the emulsion was sucked up and stirred in a beaker in which the reaction took place to build the shell of the microcapsules (Figure \ref{methods}). This process was designed with the initial aim of optimizing the size distribution of the droplets, Figure \ref{pdfR}.   However, splitting the process into two steps gave also the opportunity to minimize the dispersion of elastic properties of microcapsules based on the kinetics of shell formation.

As explained in the introduction, growth of shell based on complexation of polyelctrolytes (CH/PFacid) is limited by diffusion: $G_s$ and $h$ increase with  complexation time $t_c$ \cite{grigoriev2008new, Deniz2011,sivakumar2008monodisperse, rinaudo2008surfactant, xie2017interfacial}.  To illustrate the consequences on production by membrane emulsification, we compared two  strategies of building the shell. The naive strategy consisted in producing capsules using a concentration of PFacid of 10 \% w/w in the oil phase and collecting them for 8 min in a beaker under gentle stirring containing the same continuous phase. Microcapsules were then washed to stop the reactions (see Materials and Methods). Under these conditions, the complexation time $t_c$ was not controlled and varied between 0 and 8 min for capsules of the same batch. We expected so that capsules will have non uniform mechanical properties, because the shell of the first produced capsule should be thicker than the shell of the last produced capsule.  The black lines of Figure \ref{pdfGs_CH2}-a and b show that $P_{R}$ had a typical Gaussian  shape ($CV =$  19\%) whereas $P_{G_s}$  was broad and covered almost two orders of magnitude of $G_s$ ($CV =$ 56 \%). We explained these variations by the time lag  between production of the first and last capsules. Consequently, we designed an optimized stratagy for which $t_c$ was controlled.  Microcapsules were produced with a very low concentration of PFacid (0.1  \% w/w) in the emulsification system to stabilize the droplets and minimize the growth of the shell. Droplets were collected for 8 min in a beaker. When the production of droplets was finished, PFacid was added to the beaker to get 10 \% w/w in the oil phase and the capsules were left under gentle stirring for 8 min. In this way, $t_c$ was the same for all the microcapsules. Then $P_{G_s}$ was Gaussian  and $CV$ decreased up to 24 \%, dashed line of Figure \ref{pdfGs_CH2}-b. In conclusion, for shell formation limited by diffusion, one way to minimize the dispersion of shell elasticity  is to dissociate droplets generation and shell formation to carefully control  the reaction time. 

Growth of BSA/TC capsule is limited by the mass of proteins that is encapsulated in the droplet \cite{Clement2014,gubspun2016characterization}. As shown in our previous studies, shell growth takes a few seconds  and surface elasticity does not depend on time \cite{xie-thesis}. However, $G_s$ increases over several orders of magnitudes with the capsule size for the same physicochemical conditions \cite{Clement2014,gubspun2016characterization}. In fact, a simple mass balance shows that the surface concentration of BSA available for the reaction is proportional to the radius of the capsule \cite{Clement2014}. Then, we produced BSA droplets stabilized with a  non-ionic surfactant by membrane emulsification and collected for 8 min the droplets in a beaker with the cross-linker (TC) to build the shell. $CV$ of size distribution was about 22 \% and $P_{G_s}$ had a Gaussian shape with a mean of 0.4 N/m and a $CV$ of 25 \% (Figure \ref{BSA}). As expected from our previous studies \cite{Clement2014,gubspun2016characterization}, there was an almost linear increase with the size of the mean surface elasticty  $\overline{G_s}$ (Figure \ref{BSACHsize}). Note that  $\overline{G_s}$ did not varied in size for CH/PFacid capsules. We concluded that the size distribution has a strong impact on $P_{G_s}$ for microencapsulation systems limited by the quantity of reactants that are encapsulated in the droplet.

\begin{figure}[t]
	\centering
	\includegraphics[width=0.5\textwidth]{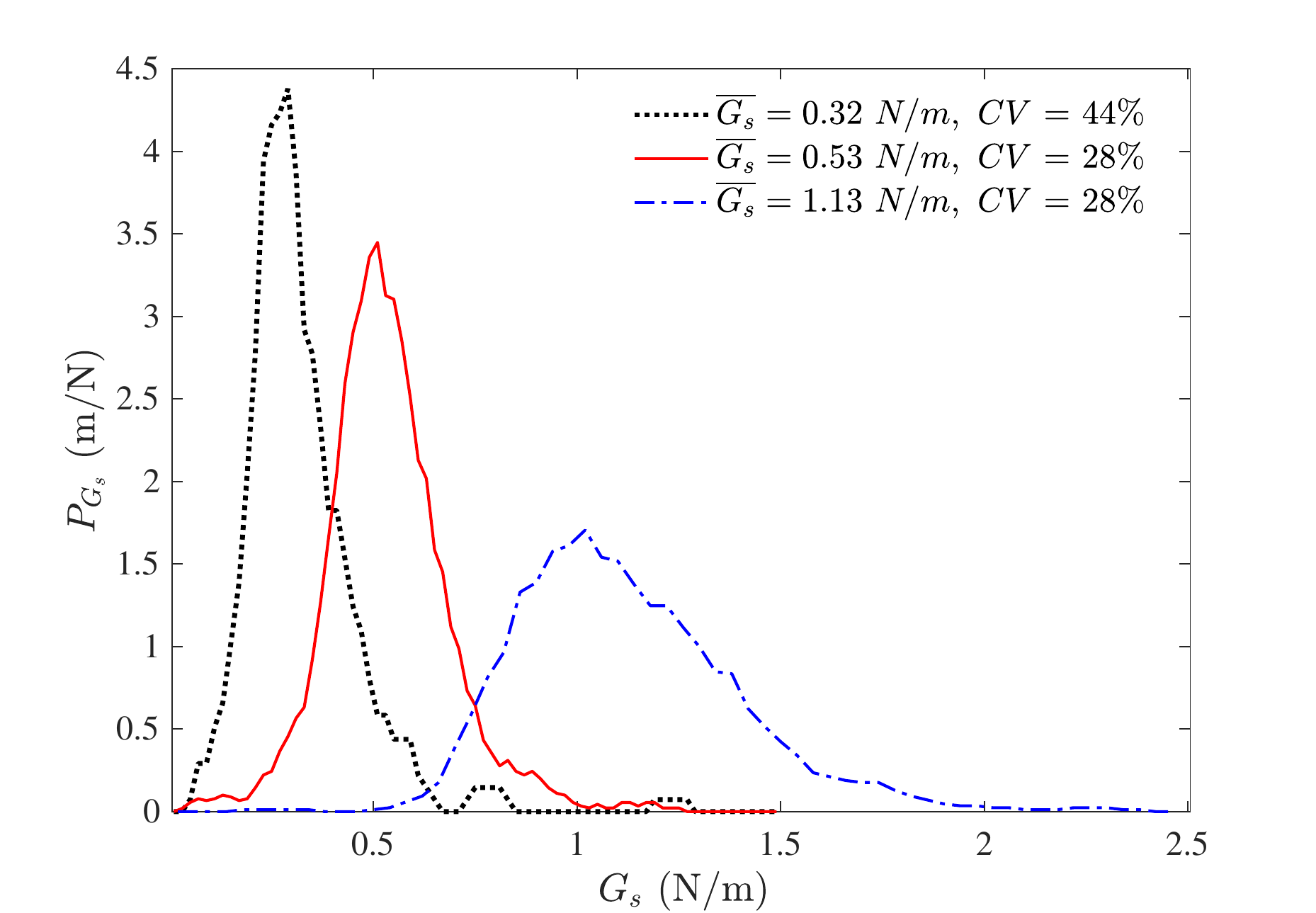}
	\caption{Probability distribution function of surface elasticity $P_{G_s}$ for different physicochemical conditions.  $\bar{G_s}$ can be tuned with the conditions. Black dotted line: PFacid 1\% w/w, CH 0.25\% w/w, $t_r =$ 4 min ; Red plain line: PFacid 1\% w/w, CH 0.25\% w/w, $t_r =$ 8 min ; Blue dashed line: PFacid21\% w/w, CH 0.25\% w/w, $t_r =$ 15 min}
	\label{pdfGs_CH}
\end{figure}

Finally,  we showed that the mean surface elasticity $\bar{G_s}$ could also be tuned by varying the concentration of chemicals in the oil and aqueous phases  or the time allowed for shell complexation after the production of the droplets (from a few seconds to severals minutes or hours). Figure \ref{pdfGs_CH} shows  $P_{G_s}$  of CH/PFacid microcapsules for various concentrations of PFacid and/or complexation times. $\overline{G_s}$  and $CV$ ranged from 0.32 to 1.13 N/m and 28 and 44 \%, respectively. Details of the effects of physicochemical conditions on shell elasticity and thickness were described in our previous publications \cite{Clement2014,xie2017interfacial,  gubspun2016characterization}.

\section{Conclusion}
A lab-scale emulsification system with micro-engineered membrane was optimized to produce suspensions of microcapsules with variations in size and elasticity of 15-20 \% in optimal conditions at throughput of a few mL/min. We concluded that the production of the droplets by membrane emulsification and capsule shell formation have to be split into two  steps when shell growth is limited by diffusion (e.g. complexation of polyelectrolytes, CH/PFacid).  If shell growth is limited by the quantity of reactants encapsulated in the droplet, variations of elastic properties are directly related to size variations (e.g. cross-linking of proteins, BSA/TC).  Finally,  membrane emulsification is a powerful method which can be easily scaled-up for the high throughput production of microcapsules with uniform and tunable physical properties, irrespective of the kinetics of shell formation.

\section{Acknowledgments}
LRP is part of the LabEx Tec21 (ANR-11-LABX-0030) and of the PolyNat Carnot Institute (ANR-11-CARN-007-01). This work has benefited from financial support from the ANR 2DVISC (ANR-18-CE06-0008).

\bibliographystyle{elsarticle-num}

\bibliography{cas-refs}

\begin{thebibliography}{10}
\expandafter\ifx\csname url\endcsname\relax
  \def\url#1{\texttt{#1}}\fi
\expandafter\ifx\csname urlprefix\endcsname\relax\def\urlprefix{URL }\fi
\expandafter\ifx\csname href\endcsname\relax
  \def\href#1#2{#2} \def\path#1{#1}\fi

\bibitem{Anandhakumar2010}
S.~Anandhakumar, V.~Nagaraja, A.~M. Raichur, Reversible polyelectrolyte
  capsules as carriers for protein delivery, Colloids Surf., B 78~(2) (2010)
  266--74.
\newblock \href {http://dx.doi.org/10.1016/j.colsurfb.2010.03.016}
  {\path{doi:10.1016/j.colsurfb.2010.03.016}}.

\bibitem{McClements2015}
D.~J. McClements, Nanoscale nutrient delivery systems for food applications:
  Improving bioactive dispersibility, stability, and bioavailability, J. Food
  Sci. 80~(7) (2015) N1602--N1611.
\newblock \href {http://dx.doi.org/10.1111/1750-3841.12919}
  {\path{doi:10.1111/1750-3841.12919}}.

\bibitem{trojer2015use}
M.~A. Trojer, L.~Nordstierna, J.~Bergek, H.~Blanck, K.~Holmberg, M.~Nyden, Use
  of microcapsules as controlled release devices for coatings, Adv. Colloid
  Interface Sci. 222 (2015) 18--43.

\bibitem{Neubauer:2014aa}
M.~P. Neubauer, M.~Poehlmann, A.~Fery, Microcapsule mechanics: from stability
  to function, Adv Colloid Interface Sci 207 (2014) 65--80.
\newblock \href {http://dx.doi.org/10.1016/j.cis.2013.11.016}
  {\path{doi:10.1016/j.cis.2013.11.016}}.

\bibitem{ChangKS1993}
K.~S. Chang, W.~L. Olbricht, Experimental studies of the deformation and
  breakup of a synthetic capsule in steady and unsteady simple shear flow, J.
  Fluid Mech. 250 (1993) 609--633.
\newblock \href {http://dx.doi.org/10.1017/s0022112093001582}
  {\path{doi:10.1017/s0022112093001582}}.

\bibitem{gao2001elasticity}
C.~Gao, E.~Donath, S.~Moya, V.~Dudnik, H.~M{\"o}hwald, Elasticity of hollow
  polyelectrolyte capsules prepared by the layer-by-layer technique, Eur. Phys.
  J. E: Soft Matter Biol. Phys. 5~(1) (2001) 21--27.

\bibitem{Clement2014}
C.~de~Loubens, J.~Deschamps, M.~Georgelin, A.~Charrier, F.~Edwards-L{\'e}vy,
  M.~Leon{\'e}tti, Mechanical characterization of cross-linked serum albumin
  microcapsules, Soft Matter 10~(25) (2014) 4561--4568.

\bibitem{xie2017interfacial}
K.~Xie, C.~De~Loubens, F.~Dubreuil, D.~Z. Gunes, M.~Jaeger, M.~L{\'e}onetti,
  Interfacial rheological properties of self-assembling biopolymer
  microcapsules, Soft matter 13~(36) (2017) 6208--6217.

\bibitem{gubspun2016characterization}
J.~Gubspun, P.-Y. Gires, C.~De~Loubens, D.~Barthes-Biesel, J.~Deschamps,
  M.~Georgelin, M.~Leonetti, E.~Leclerc, F.~Edwards-Levy, A.-V. Salsac,
  Characterization of the mechanical properties of cross-linked serum albumin
  microcapsules: effect of size and protein concentration, Colloid Polym. Sci.
  294~(8) (2016) 1381--1389.

\bibitem{richardson2016innovation}
J.~J. Richardson, J.~Cui, M.~Bj{\"o}rnmalm, J.~A. Braunger, H.~Ejima,
  F.~Caruso, Innovation in layer-by-layer assembly, Chemical Reviews 116~(23)
  (2016) 14828--14867.

\bibitem{sivakumar2008monodisperse}
S.~Sivakumar, J.~K. Gupta, N.~L. Abbott, F.~Caruso, Monodisperse emulsions
  through templating polyelectrolyte multilayer capsules, Chemistry of
  Materials 20~(6) (2008) 2063--2065.

\bibitem{hennequin2009synthesizing}
Y.~Hennequin, N.~Pannacci, C.~P. de~Torres, G.~Tetradis-Meris, S.~Chapuliot,
  E.~Bouchaud, P.~Tabeling, Synthesizing microcapsules with controlled
  geometrical and mechanical properties with microfluidic double emulsion
  technology, Langmuir 25~(14) (2009) 7857--7861.

\bibitem{kaufman2015soft}
G.~Kaufman, S.~Nejati, R.~Sarfati, R.~Boltyanskiy, M.~Loewenberg, E.~R.
  Dufresne, C.~O. Osuji, Soft microcapsules with highly plastic shells formed
  by interfacial polyelectrolyte-nanoparticle complexation, Soft matter 11~(38)
  (2015) 7478--7482.

\bibitem{de2017one}
J.~D. de~Baubigny, C.~Tr{\'e}gou{\"e}t, T.~Salez, N.~Pantoustier, P.~Perrin,
  M.~Reyssat, C.~Monteux, One-step fabrication of ph-responsive membranes and
  microcapsules through interfacial h-bond polymer complexation, Scientific
  Reports 7~(1) (2017) 1265.

\bibitem{tregouet2018microfluidic}
C.~Tregou{\"e}t, T.~Salez, C.~Monteux, M.~Reyssat, Microfluidic probing of the
  complex interfacial rheology of multilayer capsules, Soft matter 15~(13)
  (2019) 2782--2790.

\bibitem{joscelyne2000membrane}
S.~M. Joscelyne, G.~Tr{\"a}g{\aa}rdh, Membrane emulsification---a literature
  review, Journal of Membrane Science 169~(1) (2000) 107--117.

\bibitem{vladisavljevic2012production}
G.~Vladisavljevi{\'c}, I.~Kobayashi, M.~Nakajima, Production of uniform
  droplets using membrane, microchannel and microfluidic emulsification
  devices, Microfluidics and nanofluidics 13~(1) (2012) 151--178.

\bibitem{doi:10.1021/ie0504699}
S.~R. Kosvintsev, G.~Gasparini, R.~G. Holdich, I.~W. Cumming, M.~T. Stillwell,
  Liquid-liquid membrane dispersion in a stirred cell with and without
  controlled shear, Industrial \& Engineering Chemistry Research 44~(24) (2005)
  9323--9330.
\newblock \href {http://arxiv.org/abs/https://doi.org/10.1021/ie0504699}
  {\path{arXiv:https://doi.org/10.1021/ie0504699}}, \href
  {http://dx.doi.org/10.1021/ie0504699} {\path{doi:10.1021/ie0504699}}.

\bibitem{CHRISTOV200283}
N.~Christov, D.~Ganchev, N.~Vassileva, N.~Denkov, K.~Danov, P.~Kralchevsky,
  Capillary mechanisms in membrane emulsification: oil-in-water emulsions
  stabilized by tween 20 and milk proteins, Colloids and Surfaces A:
  Physicochemical and Engineering Aspects 209~(1) (2002) 83 -- 104.
\newblock \href
  {http://dx.doi.org/https://doi.org/10.1016/S0927-7757(02)00167-X}
  {\path{doi:https://doi.org/10.1016/S0927-7757(02)00167-X}}.

\bibitem{IMBROGNO2015116}
A.~Imbrogno, M.~Dragosavac, E.~Piacentini, G.~Vladisavljevi{\'c}, R.~Holdich,
  L.~Giorno, Polycaprolactone multicore-matrix particle for the simultaneous
  encapsulation of hydrophilic and hydrophobic compounds produced by membrane
  emulsification and solvent diffusion processes, Colloids and Surfaces B:
  Biointerfaces 135 (2015) 116 -- 125.
\newblock \href
  {http://dx.doi.org/https://doi.org/10.1016/j.colsurfb.2015.06.071}
  {\path{doi:https://doi.org/10.1016/j.colsurfb.2015.06.071}}.

\bibitem{VLADISAVLJEVIC201478}
G.~T. Vladisavljevi{\'c}, B.~Wang, M.~M. Dragosavac, R.~G. Holdich, Production
  of food-grade multiple emulsions with high encapsulation yield using
  oscillating membrane emulsification, Colloids and Surfaces A: Physicochemical
  and Engineering Aspects 458 (2014) 78 -- 84, formula VII: How Does Your
  Formulation Work?
\newblock \href
  {http://dx.doi.org/https://doi.org/10.1016/j.colsurfa.2014.05.011}
  {\path{doi:https://doi.org/10.1016/j.colsurfa.2014.05.011}}.

\bibitem{HANGA20141664}
M.~P. Hanga, R.~G. Holdich, Membrane emulsification for the production of
  uniform poly-n-isopropylacrylamide-coated alginate particles using internal
  gelation, Chemical Engineering Research and Design 92~(9) (2014) 1664 --
  1673.
\newblock \href {http://dx.doi.org/https://doi.org/10.1016/j.cherd.2013.12.010}
  {\path{doi:https://doi.org/10.1016/j.cherd.2013.12.010}}.

\bibitem{VLADISAVLJEVIC2014168}
G.~T. Vladisavljevi{\'c}, A.~Laouini, C.~Charcosset, H.~Fessi, H.~C.
  Bandulasena, R.~G. Holdich, Production of liposomes using microengineered
  membrane and co-flow microfluidic device, Colloids and Surfaces A:
  Physicochemical and Engineering Aspects 458 (2014) 168 -- 177, formula VII:
  How Does Your Formulation Work?
\newblock \href
  {http://dx.doi.org/https://doi.org/10.1016/j.colsurfa.2014.03.016}
  {\path{doi:https://doi.org/10.1016/j.colsurfa.2014.03.016}}.

\bibitem{fery2007mechanical}
A.~Fery, R.~Weinkamer, Mechanical properties of micro-and nanocapsules:
  Single-capsule measurements, Polymer 48~(25) (2007) 7221--7235.

\bibitem{dubreuil2003elastic}
F.~Dubreuil, N.~Elsner, A.~Fery, Elastic properties of polyelectrolyte capsules
  studied by atomic-force microscopy and ricm, Eur. Phys. J. E: Soft Matter
  Biol. Phys. 12~(2) (2003) 215--221.

\bibitem{A.Walter2000}
A.~Walter, H.~Rehage, H.Leonhard, Shear-induced deformations of polyamide
  microcapsules, Colloid Polym. Sci. 278 (2000) 169--175.

\bibitem{grigoriev2008new}
D.~Grigoriev, T.~Bukreeva, H.~M{\"o}hwald, D.~Shchukin, New method for
  fabrication of loaded micro-and nanocontainers: emulsion encapsulation by
  polyelectrolyte layer-by-layer deposition on the liquid core, Langmuir 24~(3)
  (2008) 999--1004.

\bibitem{Deniz2011}
D.~Z. Gunes, M.~Pouzot, M.~Rouvet, S.~Ulrich, R.~Mezzenga, Tuneable thickness
  barriers for composite o/w and w/o capsules, films, and their decoration with
  particles, Soft Matter 7~(19) (2011) 9206--9215.
\newblock \href {http://dx.doi.org/10.1039/c1sm05997a}
  {\path{doi:10.1039/c1sm05997a}}.

\bibitem{rinaudo2008surfactant}
M.~Rinaudo, N.~Kil’deeva, V.~Babak, Surfactant-polyelectrolyte complexes on
  the basis of chitin, Russian Journal of General Chemistry 78~(11) (2008)
  2239--2246.

\bibitem{Levy1996}
M.~C. Andry, F.~Edwards-Levy, M.~C. Levy, Free amino group content of serum
  albumin microcapsules. iii. a study at low ph values, Int. J. Appl. Pharm.
  128 (1996) 197--202.

\bibitem{lu2004microcapsule}
G.~Lu, Z.~An, C.~Tao, J.~Li, Microcapsule assembly of human serum albumin at
  the liquid/liquid interface by the pendent drop technique, Langmuir 20~(19)
  (2004) 8401--8403.

\bibitem{xie-thesis}
K.~Xie, Instabilities of microcapsules in flow: breakup and wrinkles, Ph.D.
  thesis, Centrale Marseille (2019).

\bibitem{fery2004mechanics}
A.~Fery, F.~Dubreuil, H.~M{\"o}hwald, Mechanics of artificial microcapsules,
  New journal of Physics 6~(1) (2004) 18.

\bibitem{Clement2015}
C.~de~Loubens, J.~Deschamps, G.~Boedec, M.~Leonetti, Stretching of capsules in
  an elongation flow, a route to constitutive law, J. Fluid Mech. 767 (2015)
  R3.

\bibitem{Edwards_L_vy_1993}
F.~Edwards-L{\'e}vy, M.-C. Andry, M.-C. L{\'e}vy, Determination of free amino
  group content of serum albumin microcapsules using trinitrobenzenesulfonic
  acid: effect of variations in polycondensation ph, International Journal of
  Pharmaceutics 96~(1-3) (1993) 85--90.
\newblock \href {http://dx.doi.org/10.1016/0378-5173(93)90215-2}
  {\path{doi:10.1016/0378-5173(93)90215-2}}.

\bibitem{DBB1985}
D.~Barthes-Biesel, H.~Sgaier, Role of membrane viscosity in the orientation and
  deformation of a spherical capsule suspended in shear flow, J. Fluid Mech.
  160 (1985) 119--135.

\bibitem{egidi2008membrane}
E.~Egidi, G.~Gasparini, R.~G. Holdich, G.~T. Vladisavljevi{\'c}, S.~R.
  Kosvintsev, Membrane emulsification using membranes of regular pore spacing:
  Droplet size and uniformity in the presence of surface shear, Journal of
  Membrane Science 323~(2) (2008) 414--420.

\bibitem{dripping-bertrandias}
A.~Bertrandias, H.~Duval, J.~Casalinho, M.~L. Giorgi, Dripping to jetting
  transition for cross-flowing liquids, Physics of Fluids 29~(4) (2017) 044102.
\newblock \href {http://dx.doi.org/10.1063/1.4979266}
  {\path{doi:10.1063/1.4979266}}.

\bibitem{dripping-meyer}
R.~F. Meyer, J.~C. Crocker, Universal dripping and jetting in a transverse
  shear flow, Phys. Rev. Lett. 102 (2009) 194501.
\newblock \href {http://dx.doi.org/10.1103/PhysRevLett.102.194501}
  {\path{doi:10.1103/PhysRevLett.102.194501}}.

\end{thebibliography}

\end{document}